\documentclass[review]{article}
\usepackage{lineno, graphicx, latexsym}

\newtheorem{thm}{Theorem}
\newtheorem{defin}{Definition}
\newtheorem{lemma}{Lemma}

\def\myqed{\mbox{\hspace{\fill}}\hfill $\Box$}

\title{Tree 3-spanners of diameter at most 5}

\author{
Ioannis Papoutsakis \\
Kastelli Pediados, Heraklion, Crete, Greece, 700 06
}

\begin{document}
\maketitle

\begin{abstract}
Tree spanners approximate distances within graphs; a subtree of a graph is a tree $t$-spanner of the graph
if and only if for every pair of vertices their distance in the subtree
is at most $t$ times their distance in the graph.
When a graph contains a subtree of diameter at most $t$, then trivially admits a tree
$t$-spanner. Now, determining whether a graph admits a tree
$t$-spanner of diameter at most $t+1$ is an NP complete problem, when $t\geq 4$, and it is
tractable, when $t\leq 3$. Although it is not known whether it is
tractable to decide graphs that admit a tree 3-spanner of any diameter, an efficient algorithm
to determine graphs that admit a tree 3-spanner of diameter at most 5 is presented. Moreover,
it is proved that if a graph of diameter at most 3 admits a tee 3-spanner, then it admits a tree
3-spanner of diameter at most 5. Hence, this algorithm decides tree 3-spanner admissibility of
diameter at most 3 graphs.
\end{abstract}


\section{Introduction}
There are applications of spanners in a variety of areas, such as
distributed computing \cite{Awerbuch85,Peleg89},
communication networks \cite{PelUpf88,PelResh99}, motion planning and
robotics \cite{Arikati96,Chew89}, phylogenetic analysis
\cite{Bandelt86} and in embedding finite
metric spaces in graphs approximately \cite{Rabinov98}.
In \cite{Pettie-Low-Dist-Span} it is mentioned that spanners have applications
in approximation algorithms for geometric spaces \cite{Narasimhanbook},
various approximation algorithms \cite{Fakcharoenphol}
and solving diagonally dominant linear systems \cite{Spielman}.

On one hand, in \cite{Bondy89,CaiCor95a,CaiThesis} an efficient algorithm
to decide tree 2-spanner admissible graphs is presented, where a method
to construct all the tree 2-spanners of a graph is also given. On the
other hand, in \cite{CaiCor95a,CaiThesis} it is proved that for
each $t\geq 4$ the problem to decide graphs that admit a tree $t$-spanner
is an NP-complete problem. The complexity status of the tree 3-spanner
problem is unresolved. In \cite{Fekete01}, for every $t$, an efficient algorithm to
determine whether a planar graph with bounded face length admits a tree $t$-spanner is
presented. Also, for every $t$, an efficient algorithm to
decide tree $t$-spanner admissibility of bounded degree graphs is presented in \cite{Treespannersofboundeddegree}.

Tree $t$-spanners ($t\geq 3$) have been studied for various families of
graphs. If a connected graph is a  cograph or a split graph or the
complement of a bipartite graph, then it admits a tree 3-spanner
\cite{CaiThesis}. Also, all convex bipartite graphs have a tree 3-spanner,
which can be constructed in linear time \cite{Venkatesan97}.
Efficient algorithms to recognize graphs that
admit a tree $3$-spanner have been developed for interval, permutation
and regular bipartite graphs \cite{Madanlal96}, planar graphs \cite{Fekete01},
directed path graphs \cite{Le99}, very strongly chordal graphs, 1-split graphs, and
chordal graphs of diameter at most 2 \cite{Brandstadtchordal}. This last result is
extended in this paper to diameter at most 3 general graphs, as shown in theorem~\ref{t:graphdiam}.

Moreover, every strongly chordal
graph admits a tree 4-spanner, which can be constructed in linear time
\cite{Brandst99}; note that, for each $t$, there is a connected chordal
graph that does not admit any tree $t$-spanner. In \cite{Brandstadtchordal} it is also presented
a linear time algorithm that finds a tree $t$-spanner in a small diameter chordal graph. In \cite{Manuel} the tree
$t$-spanner problem is studied for diametrically uniform graphs. An approximation algorithm for the tree
$t$-spanner problem is presented in \cite{pelegemek,DraganK}, where a new necessary condition for a graph to
have a tree t-spanner in terms of decomposition is also presented.

There are NP-completeness results for the tree $t$-spanner problem for families of graphs.
In \cite{Fekete01}, it is shown that it is NP-hard to determine the minimum $t$ for which a
planar graph admits a tree $t$-spanner. For any $t\geq4$, the tree $t$-spanner problem is NP-complete
on chordal graphs of diameter at most $t+1$, when $t$ is even, and of diameter at most $t+2$, when 
$t$ is odd \cite{Brandstadtchordal}; note that this refers to the diameter of the graph not to the diameter
of the spanner.  In \cite{Treespannersofsmalldiameter} it is shown that the
problem to determine whether a graph admits a tree $t$-spanner of
diameter at most $t+1$ is tractable, when $t\leq 3$, while it is an
NP-complete problem, when $t\geq 4$. For example, deciding graphs that admit a tree 4-spanner of diameter
at most 5 is an NP-complete problem. In this paper, an efficient algorithm to decide graphs that admit
a tree 3-spanner of diameter at most 5 is presented (theorem~\ref{t:final}).

The tree 3-spanner problem is very interesting, since its complexity status is unresolved. In \cite{PhDthesis} it is
shown that only for $t=3$ the union of any two tree $t$-spanners of any given graph may contain big induced cycles but never
an odd induced cycle (other than a triangle); such unions are proved to be perfect graphs. The algorithm presented in
\cite{Treespannersofboundeddegree} is efficient only for $t\leq3$, when graphs with maximum degree $O(\log n)$
are considered, where $n$ is the number of vertices of each graph.
The tree 3-spanner problem can be formulated as an integer
programming optimization problem. Constraints for such a formulation appear in \cite{PhDthesis}, providing certificates
of tree 3-spanner inadmissibility for some graphs.

\section{Definitions and lemmas}
In general, terminology of \cite{West} is used. Let $G$ be a graph. Then, $V(G)$ is its vertex set and
$E(G)$ its edge set. An edge between vertices $u,v\in G$ is denoted as $uv$.
Let $v$ be a vertex of $G$, then $N_G(v)$ is the set of $G$ neighbors of $v$, while
$N_G[v]$ is $N_G(v)\cup \{v\}$; in this paper we consider graphs without loop edges, so $v\not\in N_G(v)$.
The closed and open neighborhoods of a subgraph $H$ of $G$ are defined as follows: $N_G[H]=\bigcup_{x\in V(H)}N_G[x]$
and $N_G(H)=N_G[H]\setminus V(H)$.
If $H$ is a subgraph of $G$, then $G[H]$ is the subgraph of $G$ induced by the vertices
of $H$, i.e. $G[H]$ contains all the vertices of $H$ and all the edges of $G$ between vertices of $H$.
The $G$ distance between two connected in $G$ to each other vertices $u,v$ is the length of a
$u, v$ shortest path in $G$ and it is denoted as $d_G(u,v)$. The diameter of a graph is the maximum distance among
pairs of vertices of the graph. The components of $G$ are its maximal connected
subgraphs. Also, a block of $G$ is a maximal connected subgraph of $G$ with no cut-vertex, where a cut-vertex is a
vertex whose deletion increases the number of components of $G$. 

Let $f, g$ be functions from the set of all graphs to the non negative integers. Then, $f$ is $O(g)$ if and only
if there are graph $G_0$ and integer $M$ such that $f(G)\leq M g(G)$ for every $G$ with $|V(G)|>|V(G_0)|$.
The definition of a tree $t$-spanner follows.

\begin{defin}
A graph $T$ is a tree $t$-spanner of a graph $G$ if and only if $T$ is a
subgraph of $G$ that is a tree and, for every pair $u$ and $v$ of vertices of
$G$, if $u$ and $v$ are at distance $d$ from each other in $G$, then $u$
and $v$ are at distance at most $t\cdot d$ from each other in $T$.
\end{defin}

Finding the minimum $t$ for which a given graph admits a tree $t$-spanner is known as
the minimum max-stretch spanning tree problem.
Note that in order to check whether a spanning tree of a graph $G$
is a tree $t$-spanner of $G$, it suffices to examine pairs of adjacent
in $G$ vertices. To focus on trees that have diameter at most 5, the concept of a 5-center
is introduced.
\begin{defin}
\label{d:center}
A 5-center of a tree $T$ consists of a pair of adjacent in $T$ vertices $u$ and $v$, such that all
vertices of $T$ are within distance 2 from $u$ or $v$ in $T$.
\end{defin}

Clearly, if a tree admits a 5-center, then it has diameter at most 5. Also, if a tree
has diameter at most 5 and contains at least one edge, then it admits a 5-center.
A frequently used lemma follows.
\begin{lemma}
Let $G$ be a graph and $T$ a tree 3-spanner of $G$. If $u$ is in a
$p,\,q$-path of $T$ and $p$, $q$ are not in $N_T[u]$,
then every $p,\,q$-path of $G$ contains a vertex in $N_T[u]$.
\label{l:cutset}
\end{lemma}
{\it Proof}.
Consider the components of $T\setminus u$. Obviously, vertices $p$ and
$q$ belong to different such components. Therefore, for any $p,\,q$-path
$P'$ of $G$ there is an edge $ww'$ in $P'$ such that $w$ and $w'$ are also in different
such components.
Since all the tree paths connecting vertices of different such components
pass through $u$, it holds that $d_T(w,w')=d_T(w,u)+d_T(u,w')$.
But the tree distance between $w$ and $w'$ can be at most
3; therefore, at least one of $w$ or $w'$ is at distance at most 1 from $u$ in $T$.
\myqed

It turns out that it suffices to examine tree 3-spanners with a 5-center
whose vertices are as close to the 5-center as possible.

\begin{defin}
\label{d:concentrated}
A tree 3-spanner $T$ of a graph $G$ is $uv$-concentrated
if and only if all of the following hold:
\begin{enumerate}
\item pair $u$, $v$ is a 5-center of $T$,
\item all $G$ neighbors of $u$ that are closer in $T$ to $u$ than to $v$  are also $T$ neighbors of $u$, and
\item all $G$ neighbors of $v$ that are closer in $T$ to $v$ than to $u$  are also $T$ neighbors of $v$.
\end{enumerate}
\end{defin}

\begin{lemma}
\label{l:concentrated}
If $G$ admits a tree 3-spanner with 5-center $uv$, then $G$ admits a
$uv$-concentrated tree 3-spanner.
\end{lemma}
{\it Proof.} Let $T$ be a tree 3-spanner of $G$ with 5-center $uv$ and let
$w$ be a vertex of $T$ which certifies that $T$ is not $uv$-concentrated.
Without loss of generality, assume that $w$ is a $G$ neighbor of
$u$, that $w$ is closer in $T$ to $u$ than to $v$, and that $w$ is not a $T$ neighbor of $u$.
Hence, $w$ is a leaf of $T$. Let $q$ be the $T$ neighbor of $w$.
Then, graph $T'$ with vertex set $V(T)$ and edge set $(E(T)\setminus\{wq\})\cup\{wu\}$ is a tree
3-spanner of $G$ with 5-center $uv$. But $T'$ has fewer than $T$ vertices
which certify that $T'$ is not $uv$-concentrated. \myqed

\section{Description of the algorithm}
\label{s:description}

\begin{figure}[htbp]
\begin{center}
\begin{tt}
\fbox{
\parbox{0.8\linewidth}{
{\bf Algorithm T3SD5}($G$)\newline
{\bf Input}. A graph $G$.\newline

{\bf If} ($E(G)=\emptyset$)\newline
\mbox{\hspace{.5cm}}{\bf If} ($|V(G)|\leq 1$) {\bf return} YES; {\bf else return} NO\hfill /*(1)\newline
{\bf For} (edge $uv$ in $G$)\{\newline
\mbox{\hspace{.5cm}}${\cal Q} =\{Q\subseteq G: Q \mbox{ is a component of } G\setminus N_G[u,v]\}$\newline
\mbox{\hspace{.5cm}}{\bf If} (${\cal Q}=\emptyset$) {\bf return} YES\hfill /*(2)\newline
\mbox{\hspace{.5cm}}${\cal C}^u=\emptyset$; ${\cal C}^v=\emptyset$\newline
\mbox{\hspace{.5cm}}{\bf do}\{\newline
\mbox{\hspace{1cm}}{\bf Pick} $Q\in{\cal Q}\setminus\bigcup_{X\in{\cal C}^u}X.M$\newline
\mbox{\hspace{1cm}}${\cal C}^u={\cal C}^u\cup\{${\bf Get\_structure}($G, u, {\cal Q}, Q)\}$\newline
\mbox{\hspace{.5cm}}\}{\bf while} (${\cal Q}\supset\bigcup_{X\in{\cal C}^u}X.M$)\newline
\mbox{\hspace{.5cm}}{\bf do}\{\newline
\mbox{\hspace{1cm}}{\bf Pick} $Q\in{\cal Q}\setminus\bigcup_{X\in{\cal C}^v}X.M$\newline
\mbox{\hspace{1cm}}${\cal C}^v={\cal C}^v\cup \{${\bf Get\_structure}($G, v, {\cal Q}, Q)\}$\newline
\mbox{\hspace{.5cm}}\}{\bf while} (${\cal Q}\supset\bigcup_{X\in{\cal C}^v}X.M$)\newline
\mbox{\hspace{.5cm}}$\Gamma=$ {\bf Create\_graph}(${\cal C}^u, {\cal C}^v$)\newline
\mbox{\hspace{.5cm}}{\bf If} ({\bf Check\_clique}($\Gamma, {\cal C}^u, {\cal C}^v, {\cal Q}$)) {\bf return} YES\hfill /*(3)\newline
\}\newline
{\bf return} NO
}}
\end{tt}
\caption{Algorithm {\tt T3SD5($G$)} that decides whether $G$ admits a tree 3-spanner of diameter at most 5. Note that
immediately after executing a {\tt return} command the algorithm halts.}
\label{f:algorithm}
\end{center}
\end{figure}

An algorithm to decide whether a given graph $G$ admits a tree 3-spanner of diameter at most 5 is described. The main function
of this algorithm is named {\tt  T3SD5}, appears in figure~\ref{f:algorithm}, and calls other functions that appear in succeeding figures. If $G$ has no edges, then $G$ suffices to be connected in order to admit a tree spanner. So, after handling this trivial
case, algorithm {\tt  T3SD5} starts examining each edge of $G$, since each edge may be the central edge of
an anticipated tree 3-spanner of $G$. So, given an edge $uv$ of $G$, the components of $G\setminus N_G[u,v]$ are stored
in set ${\cal Q}$. Of course, if  ${\cal Q}$ is empty, then $G$ immediately admits a small diameter (at most 3) tree 3-spanner.
The key idea of this algorithm is to examine each member of  ${\cal Q}$. Using lemma~\ref{l:cutset}, one can prove
that each component in  ${\cal Q}$ is completely placed on one side (towards $u$ or towards $v$) of a tree 3-spanner of $G$
with central edge $uv$.
 
\begin{figure}[htbp]
\begin{center}
\begin{tt}
\fbox{
\parbox{0.8\linewidth}{
{\bf Function Get\_structure}($G, x, {\cal Q}, Q)$\newline
{\bf Input}. A graph $G$, a vertex $x$, a set of components ${\cal Q}$,
and a component $Q$.\newline

{\bf New structure} $C$\newline
Let $Q^x$ be the component of $G\setminus N_G[x]$ that contains $Q$.\newline
$C.M=\{X\in{\cal Q}: X\subseteq Q^x\}$\newline
$C.U=Q^x\setminus\bigcup C.M$\newline
$C.D=N_G(Q^x)$\newline
$C.R=\{z\in C.D: N_G(z)\supseteq Q^x\}$\newline
{\bf return} $C$
}}
\end{tt}
\caption{Function {\tt Get\_structure($G, x, {\cal Q}, Q)$}.}
\label{f:structures}
\end{center}
\end{figure}

Consider the case of a tree 3-spanner $T$ of $G$ for which a component $Q$ in ${\cal Q}$ is placed on the side of $u$, for
example. Then, some other components
must follow $Q$ and also some vertices in $N_G(u,v)$ must be placed on the same side as $Q$. To collect these implications in an
orderly manner, given $Q$ and $u$, a structure $C$ is formed by calling function {\tt Get\_structure} in figure~\ref{f:structures}.
There, $Q^x$ is the component of $G\setminus N_G[x]$ that contains $Q$, where $x=u$ in this call of function {\tt Get\_structure}.
First, all components of ${\cal Q}$ that are in $Q^x$ must follow $Q$ and are placed in set $C.M$ of structure $C$; the
coMponents of $C$. Second, the remaining vertices of $Q^x$ (these are neighbors of $v$) must also follow $Q$. These
vertices will be at distance 2 from $u$ in $T$ (because they are not neighbors of $u$) and are stored in
set $C.U$; the Up vertices of $C$. Third the neighbors of $Q^x$ must follow $Q$ again. They can be at distance 1 from $u$ in
$T$ (because they are neighbors of $u$) and are stored in set $C.D$; the Down vertices of $C$.
Finally, fourth, it can be proved that all vertices in $Q^x$ are adjacent in $T$ to one vertex in $C.D$. Set $C.R$ stores
all such candidates; the Representatives of $C$. 

\begin{figure}[htbp]
\begin{center}
\begin{tt}
\fbox{
\parbox{0.8\linewidth}{
{\bf Function Create\_graph}(${\cal C}^u, {\cal C}^v$)\newline
{\bf Input}. Two disjoint sets of structures.\newline

$V=\{X\in {\cal C}^u\cup{\cal C}^v: X.R\not=\emptyset\}$\newline
$E=\{XY\in V^2: (X\in {\cal C}^u$ AND $Y\in {\cal C}^u)$ OR\newline
\mbox{\hspace{.5cm}}$(X\in {\cal C}^v$ AND $Y\in {\cal C}^v)$ OR\newline
\mbox{\hspace{.5cm}}$(X\in {\cal C}^u$ AND $Y\in {\cal C}^v$ AND $X.M\cap Y.M=\emptyset$ AND\newline
\mbox{\hspace{1.5cm}}$(X.U\cup X.D)\cap(Y.U\cup Y.D)=\emptyset)\}$\newline
{\bf return} $(V,E)$
}}
\end{tt}
\caption{Function {\tt Create\_graph(${\cal C}^u, {\cal C}^v$)}.}
\label{f:graph}
\end{center}
\end{figure}

All these structures are placed in sets ${\cal C}^u$ and ${\cal C}^v$. The aim is to tile these structures of implications in
a way that a tree 3-spanner is formed if possible. Towards this aim, a graph $\Gamma$ is formed based on these structures
by calling function {\tt Create\_graph} in figure~\ref{f:graph}. Each structure becomes a vertex of $\Gamma$ and edges
are placed between compatible structures. Note that $\Gamma$ is the complement of a bipartite graph. To achieve this aim,
graph $\Gamma$ must contain some clique of structures that covers all the components in ${\cal Q}$; this is decided by
calling function {\tt Check\_clique} in figure~\ref{f:clique}. There, if a component in ${\cal Q}$ belongs only in one
structure, then this structure must be in the clique and all its non neighbors must not. Finally, having such a clique of $\Gamma$,
one can find a tree 3-spanner of $G$ of diameter at most 5, using function {\tt FT3SD5} in figure~\ref{f:function}. Note
that the clique suggested here produces a spanner that has as many components of ${\cal Q}$ on the side of $u$ as possible.

\begin{figure}[htbp]
\begin{center}
\begin{tt}
\fbox{
\parbox{0.8\linewidth}{
{\bf Function Check\_clique}($\Gamma, {\cal C}^u, {\cal C}^v, {\cal Q}$)\newline
{\bf Input}. A graph $\Gamma$ and two disjoint sets ${\cal C}^u$ and ${\cal C}^v$ that cover its vertex set.\newline

$K=\emptyset$; $V=V(\Gamma)$\newline
flag = (${\cal Q}\subseteq\bigcup_{X\in V}X.M$)\newline
{\bf while} (${\cal Q}\supset(\bigcup_{X\in{\cal C}^u\cap V}X.M)\cup(\bigcup_{X\in K}X.M)$ AND flag)\{\newline
\mbox{\hspace{.5cm}}$K=K\cup\{X\in{\cal C}^v\cap V:$ there exists $Q\in X.M$\newline
\mbox{\hspace{1.5cm}}such that $Q\not\in\bigcup_{Y\in{\cal C}^u\cap V}Y.M$\}\hfill /*(1)\newline
\mbox{\hspace{.5cm}}$V=V\setminus\{X\in V:$ there exists $Y\in K$\newline
\mbox{\hspace{1.5cm}}such that $XY\not\in E(\Gamma)\}$\newline
\mbox{\hspace{.5cm}}flag = (${\cal Q}\subseteq\bigcup_{X\in V}X.M$)\newline
\mbox{\hspace{.5cm}}\}\newline
{\bf If} (flag) {\bf return} 1; {\bf else return} 0
}}
\end{tt}
\caption{Function {\tt Check\_clique($\Gamma, {\cal C}^u, {\cal C}^v, {\cal Q}$)}.}
\label{f:clique}
\end{center}
\end{figure}

\section{Proof of correctness}
\begin{lemma}
Assume that algorithm {\tt T3SD5} in figure~\ref{f:algorithm} is run on input a graph $G$ and that edge $uv$ is
examined in its {\tt for} loop. Also, assume that set ${\cal Q}$ formed upon $uv$ is not empty.
Then\footnote{Note that the conclusion of the lemma holds for $v$ as well: for every component
$W\in{\cal Q}$, there exists a unique structure $C\in{\cal C}^v$, such that $W\in C.M$.}, for every component
$W\in{\cal Q}$, there exists a unique structure $C\in{\cal C}^u$, such that $W\in C.M$.
\label{l:partition}
\end{lemma}
{\em Proof}. Since ${\cal Q}$ is not empty the algorithm proceeds with the construction of set of
structures ${\cal C}^u$ through its first {\tt do-while} loop.
At each step of this construction, function {\tt Get\_structure} (figure~\ref{f:structures}) is called and a subset of ${\cal Q}$ is placed in the returned structure; also, this structure is added to ${\cal C}^u$. The construction proceeds to the next step,
until all components of ${\cal Q}$ are placed in various structures. The {\tt do-while} loop of this construction terminates,
because,first, at each step at least one non placed yet component is placed and, second, function {\tt Get\_structure}
always returns. So, there exists a structure $C\in{\cal C}^u$, such that $W\in C.M$.

To form $C$ function {\tt Get\_structure} is called with input ($G, u, {\cal Q}, Q$), where
$Q$ is some component in ${\cal Q}$. Assume that there is another structure $C'\in{\cal C}^u$, such that $W\in C'.M$.
Again, $C'$ must be formed by calling function {\tt Get\_structure} with input ($G, u, {\cal Q}, Q'$), where
$Q'$ is some component in ${\cal Q}$. Both of $Q$ and $Q'$ must be in the component of $G\setminus N_G[u]$ that
contains $W$, so $Q\in C'.M$ and $Q'\in C.M$. Without loss of generality, assume that structure $C$ is
formed first. Since $Q'\in C.M$, algorithm {\tt T3SD5} cannot pick $Q'$ in order to call function {\tt Get\_structure}
with input ($G, u, {\cal Q}, Q'$), a contradiction.\myqed

\begin{lemma}
\label{l:necessary}
Let $G$ be a graph that admits a $uv$-concentrated tree 3-spanner $T$.
Let $W$ be a component of $G\setminus N_G[u, v]$. Assume that
$W$ contains a vertex which is at distance 2 from $u$ in $T$. Then, algorithm {\tt T3SD5} in figure~\ref{f:algorithm}
on input $G$ returns {\tt YES} or the following hold:
\begin{enumerate}
\item There exists a structure $C\in{\cal C}^u$, such that $W\in C.M$, where ${\cal C}^u$ is the set
of structures constructed by algorithm {\tt T3SD5} on input $G$, when edge $uv$ is
examined in its {\tt for} loop.
\item There exists an $r\in C.D$, such that every vertex in $C.U\cup\bigcup C.M$ is adjacent to $r$ in $T$.
\item Every vertex in $C.D$ is adjacent to $u$ in $T$.
\end{enumerate}
\end{lemma}
{\em Proof}. Assume that algorithm {\tt T3SD5} is run on input $G$. If it has not not returned {\tt YES},
edge $uv$ of $G$ is examined in its {\tt for} loop. So, set ${\cal Q}$ is formed based on $u$ and $v$.
Here, ${\cal Q}$ contains at least one component, namely $W$, so it is not empty. So, by lemma~\ref{l:partition}
there exists a structure $C\in{\cal C}^u$, such that $W\in C.M$.

Let $p$ be the vertex of $W$ which it is known to be at distance 2 from $u$ in $T$. Then, there is a vertex
$r$, such that $pr$ and $ru$ are edges of $T$. Assume that there is a vertex $q$ in $X.U\cup\bigcup X.M$
that is not adjacent to $r$ in $T$. There is a path from $q$ to $u$ in $T$. This path avoids $p$, because
$p$ is a leaf ($u, v$ is a 5-center of $T$ and $p$ is not in $N_G[u, v]$). It also avoids $r$,
because all $T$ neighbors of $r$ but $u$ are leaves ($u, v$ is a 5-center of $T$ and $r$ is adjacent to $u$ in $T$)
different than $q$. So, $u$ is in the tree path from $p$ to $q$ and $u$ is not a $T$ neighbor of either $p$ or $q$
(note that $q\not\in N_G[u]$).
There is a path from $p$ to $q$ in $G\setminus N_G[u]$, because $G[C.U\cup\bigcup C.M]=Q^x$
(figure~\ref{f:structures}; note that $x=u$ here) and $Q^x$ is a connected graph that doesn't overlap with
$N_G[u]$. But this is a contradiction to lemma~\ref{l:cutset}. Therefore, every vertex in $C.U\cup\bigcup C.M$ is
adjacent to $r$ in $T$. Clearly, $p$ belongs to $Q^x$ and $r$ doesn't belong to $Q^x$. So, $r$ belongs to
$N_G(Q^x)$ and therefore belongs to $C.D$.

Let $w$ be in $C.D$. All $T$ neighbors of $r$ but $u$ are leaves and
every vertex in $C.U\cup\bigcup C.M=Q^x$ is adjacent to $r$ in $T$.  So, $w$ must be within distance
2 from $r$, because $w$ is a $G$ neighbor of a vertex in $Q^x$ and $T$ is a 3-spanner of
$G$. Here, $w$ cannot be adjacent to $r$, because $T$ is $uv$-concentrated. The $T$ path of length 2 from
$w$ to $r$ must contain $u$, because $u$ is the only non leaf neighbor of $r$. This makes $w$ adjacent to $u$ in
$T$.\myqed

\begin{lemma}
Let ${\cal Q}$ be the set of components of $G\setminus N_G[u, v]$, where $G$ is a graph, $uv$ an edge of $G$ and
${\cal Q}$ is not empty. Also, let ${\cal C}^u$ and ${\cal C}^v$ be the sets of structures formed when edge $uv$ is
examined by algorithm  {\tt T3SD5} in figure~\ref{f:algorithm} on input $G$. Finally, let $\Gamma$ be the graph
constructed by the algorithm upon ${\cal C}^u$ and ${\cal C}^v$. Then, function {\tt Check\_clique} in
figure~\ref{f:clique} on input ($\Gamma, {\cal C}^u, {\cal C}^v, {\cal Q}$) returns 1 if and only if
$\Gamma$ contains a clique $L$ such that $\bigcup_{X\in L}X.M={\cal Q}$.
\label{l:clique}
\end{lemma}

{\em Proof}. Assume that $\Gamma$ contains a clique $L$ such that $\bigcup_{X\in L}X.M={\cal Q}$.
Whenever the conditions of the {\tt while} loop of function {\tt Check\_clique} are checked,
$K\subseteq L\subseteq V$; this is proved by induction on the number of times these
conditions are checked. Let the base case be the first time the {\tt while} statement is executed.
At this point $K=\emptyset$ and $V=V(\Gamma)$, so $K\subseteq L\subseteq V(\Gamma)$.
For the induction step, first, each $X$ added to $K$ is the only vertex in $V$ that contains (in its X.M) a specific
component $Q$ of ${\cal Q}$, because $Q$ is not contained in any vertex in ${\cal C}^u\cap V$ (see condition (1)
in figure~\ref{f:clique}) and $Q$ is contained in a unique structure-vertex of ${\cal C}^v$ (lemma~\ref{l:partition}).
So, $X\in L$, because $\bigcup_{X\in L}X.M={\cal Q}$ and $L\subseteq V$ (from the induction hypothesis).
Therefore, $K\subseteq L$.
Second, from $V$ are removed all the vertices that are not adjacent to at least one vertex of $K$, but
none of these vertices can be in $L$, since $K\subseteq L$ and $L$ is a clique. So, $L\subseteq V$.

Boolean variable {\tt flag} remains equal to 1 during the execution of the {\tt while} statement of function
{\tt Check\_clique}, because $\bigcup_{X\in L}X.M={\cal Q}$ and $L\subseteq V$. If at some step of
the {\tt while} execution set $K$ is not increased, then $\bigcup_{X\in({\cal C}^v\cap V)\setminus K} X.M
\subseteq\bigcup_{X\in{\cal C}^v\cap V} X.M$. But $V\subseteq{\cal C}^v\cup{\cal C}^u$, by construction
of graph $\Gamma$ (figure~\ref{f:graph}), and $Q\subseteq\bigcup_{X\in V}X.M$, because {\tt flag} is 1. So,
 ${\cal Q}\subseteq(\bigcup_{X\in{\cal C}^u\cap V}X.M)\cup(\bigcup_{X\in K}X.M)$ and the {\tt while} loop
terminates. But $K$ can't increase for ever, because it is bounded by ${\cal C}^v\cap V$. So, the {\tt while} loop
does terminate and, of course, the function returns 1.

Assume that function {\tt Check\_clique} returns 1. Then, {\tt flag} is equal to 1 and the {\tt while} statement
terminates. So, $Q\subseteq(\bigcup_{X\in{\cal C}^u\cap V}X.M)\cup(\bigcup_{X\in K}X.M)$. Set $L$ equal to
$({\cal C}^u\cap V)\cup K$. But for every $X\in V(\Gamma)$ it holds that $X.M\subseteq{\cal Q}$. So,
$\bigcup_{X\in L}X.M={\cal Q}$.

First, $K\subseteq{\cal C}^v\cap V(\Gamma)$ (see formation of set $K$ in command (1) in figure~\ref{f:clique}). Vertices
in ${\cal C}^v\cap V(\Gamma)$ form a clique, because of definition of edge set of $\Gamma$ in figure~\ref{f:graph}.
So, $K$ forms a clique in $\Gamma$. Second, there is no vertex in $V$ which is not adjacent to all vertices in $K$.
Therefore $L$ forms a clique in $\Gamma$.\myqed

\begin{thm}
\label{t:final}
A graph $G$ admits a tree 3-spanner of diameter at most 5 if and only
if algorithm {\tt T3SD5} in figure~\ref{f:algorithm} on input $G$ returns {\tt YES}.
\end{thm}
{\em Proof}. On one hand assume that a graph $G$ admits a tree 3-spanner $T'$ of diameter at most 5. If $G$ has no
edges, then $G$ must have at most one vertex. Then, in this case, algorithm {\tt T3SD5} on input $G$
returns {\tt YES}. So, assume that $G$ has at least one edge. Then, $T'$ contains
at least one edge too. So, since $T'$ has diameter at most 5, $T'$ contains two adjacent vertices
$u$ and $v$ that form a 5-center of $T'$. Therefore, by lemma~\ref{l:concentrated}, $G$ admits
a $uv$-concentrated tree 3-spanner $T$.

Algorithm {\tt T3SD5} on input $G$ starts examining edges through its {\tt for} loop. If the algorithm has not
returned {\tt YES} yet, it examines edge $uv$. Then, set of components ${\cal Q}$ is formed. If ${\cal Q}$ is empty,
the algorithm returns {\tt YES}. So, assume that ${\cal Q}$ is not empty. Therefore, sets ${\cal C}^u$ and ${\cal C}^v$
are constructed and upon these sets graph $\Gamma$ is built. Let $A$ be the subset of ${\cal C}^u$ such that
$X\in A$ if and only if a vertex in a component in $X.M$ is at distance 2 from $u$ in $T$. So, according to
lemma~\ref{l:necessary} (conclusion 2), for every $X\in A$,  there exists an $r\in X.D$, such that every edge from $r$ to
$X.U\cup V(\bigcup X.M)$ belongs to $T$. So, for every $X\in A$, $X.R\not=\emptyset$. Therefore, $A\subseteq V(\Gamma)$
(vertex set formation of function {\tt Create\_graph} in figure~\ref{f:graph}). Similarly, $B\subseteq V(\Gamma)$,
where $B$ is the subset of ${\cal C}^v$ such that $X\in B$ if and only if  a vertex in a component in $X.M$ is at distance
2 from $v$ in $T$.

First, graph $\Gamma$ contains all the edges between its vertices that belong to ${\cal C}^u$ (edge set in
function {\tt Create\_graph}).
Since all structures in $A$ belong to ${\cal C}^u$, $A$ forms a clique in $\Gamma$; similarly $B$ forms a clique in $\Gamma$.
Second, let $X\in A$ and $Y\in B$. All vertices in $X.U\cup V(\bigcup X.M)$ are adjacent in $T$ to a vertex in $X.D$ and
all vertices in $X.D$ are adjacent to $u$ in $T$, because of conclusions 2 and 3 of lemma~\ref{l:necessary}.
So, every vertex in $X.D\cup X.U\cup V(\bigcup X.M)$ is connected to
$u$ in $T$ through a path that avoids $v$. Similarly, every vertex in $Y.D\cup Y.U\cup V(\bigcup Y.M)$ is connected to
$v$ in $T$ through a path that avoids $u$. So, sets $X.D\cup X.U\cup V(\bigcup X.M)$ and $Y.D\cup Y.U\cup V(\bigcup Y.M)$
must be disjoint, because otherwise $T$ would contain a cycle, since $uv\in T$. Therefore, edge $XY$ is in $\Gamma$ (see
definition of edge set in function {\tt Create\_graph}). From these two facts, $A\cup B$ forms a clique in $\Gamma$.

Every component $Q$ in ${\cal Q}$ contains a vertex at $T$ distance 2 from $u$ or $v$, because $Q\cap N_G[u, v]=\emptyset$
and $u, v$ form a 5-center of $T$. So, there is an $X$ in $A\cup B$ such that $Q\in X.M$, because of the definitions of
$A$ and $B$ and lemma~\ref{l:partition}. Therefore, $A\cup B$ is a clique of $\Gamma$ that covers all
the components in ${\cal Q}$. Then, by lemma~\ref{l:clique}, function {\tt Check\_clique} on input
($\Gamma, {\cal C}^u, {\cal C}^v, {\cal Q}$) returns 1 and, therefore, algorithm {\tt T3SD5} on input $G$ returns {\tt YES}.

On the other hand assume that algorithm {\tt T3SD5} on input $G$ returns {\tt YES}. The algorithm returns {\tt YES} in 3
cases. First, command (1) in figure~\ref{f:algorithm}. Then, $G$ has no edges and at most one vertex, so it trivially
admits a tree 3-spanner of diameter at most 5. Second, command (2), while examining some edge $uv$ of $G$. Then,
{\cal Q} is empty. This means that $N_G[u,v]$ covers $G$. In this case let $T$ be the spanning tree of $G$ with edge
set $\{ux: x\in N_G(u)\}\cup\{vx: x\in N_G(v)\setminus N_G[u]\}$. Then, $T$ is a tree 3-spanner of diameter at most 3.

\begin{figure}[htbp]
\begin{center}
\begin{tt}
\fbox{
\parbox{0.8\linewidth}{
{\bf Function FT3SD5}($G, {\cal C}^u, {\cal C}^v, L$)\newline
{\bf Input}. A graph $G$ and appropriate sets of structures based on $G$.\newline

$V=V(G)$\newline
$E_U=\emptyset$; $E_D=\emptyset$\newline
{\bf For} (structure $X$ in $L$)\{\newline
\mbox{\hspace{.5cm}}{\bf Pick} $r\in X.R$\newline
\mbox{\hspace{.5cm}}$E_U=E_U\cup\{rx: x\in X.U\cup\bigcup X.M\}$\newline
\mbox{\hspace{.5cm}}{\bf If} ($X\in{\cal C}^u$)\newline
\mbox{\hspace{1cm}}$E_D=E_D\cup\{ux: x\in X.D\}$\newline
\mbox{\hspace{.5cm}}{\bf If} ($X\in{\cal C}^v$)\newline
\mbox{\hspace{1cm}}$E_D=E_D\cup\{vx: x\in X.D\}$\newline
\}\newline
$V_L=\bigcup_{X\in L}(X.U\cup X.D)$\newline
$E'=\{ux: x\in N_G(u)\setminus V_L\}\cup\{vx: x\in N_G(v)\setminus (N_G[u]\cup V_L)\}$\newline
$E=E'\cup E_U\cup E_D$\newline
{\bf return} $(V, E)$\newline
}}
\end{tt}
\caption{Function {\tt FT3SD5($G, {\cal C}^u, {\cal C}^v, L$)} that returns a tree 3-spanner of $G$ of diameter at most 5.
Here, $L$ can be equal to $({\cal C}^u\cap V)\cup K$, where sets $V$ and $K$ are constructed in function {\tt Check\_clique}}
\label{f:function}
\end{center}
\end{figure}

Third, command (3) of algorithm {\tt T3SD5}. Then, for some edge $uv$ of $G$ function {\tt Check\_clique} in
figure~\ref{f:clique} on input ($\Gamma, {\cal C}^u, {\cal C}^v, {\cal Q}$) returns 1, where
$\Gamma, {\cal C}^u, {\cal C}^v, {\cal Q}$ are constructed by the algorithm upon edge $uv$. So, by lemma~\ref{l:clique},
$\Gamma$ contains a clique $L$ such that $\bigcup_{X\in L}X.M={\cal Q}$. In the proof of this lemma some such $L$ is
presented and it is equal to $({\cal C}^u\cap V)\cup K$, where sets $V$ and $K$ are constructed in function {\tt Check\_clique}.
Given this $L$, let $T$ be the graph returned by function {\tt FT3SD5} of figure~\ref{f:function} on input
($G, {\cal C}^u, {\cal C}^v, L$).

For each $Q\in{\cal Q}$ there is a unique $X\in L$ such that $Q\in X.M$. To see this, first, $\bigcup_{X\in L}X.M={\cal Q}$, so
there is at least one $X\in L$ such that $Q\in X.M$. Second, assume that there is another $Y\in L$, such that
$Q\in Y.M$. If both $X$ and $Y$ belong to ${\cal C}^u$, then this is a contradiction to lemma~\ref{l:partition}; the same holds
if they both belong to ${\cal C}^v$. Without loss of generality, assume that $X\in{\cal C}^u$ and $Y\in{\cal C}^v$. But $L$
is a clique of $\Gamma$, so edge $XY$ is in $\Gamma$. By construction of edge set of $\Gamma$ (figure~\ref{f:graph}),
$X.M\cap Y.M=\emptyset$, a contradiction. Therefore, each vertex in $\bigcup{\cal Q}$ has degree 1, because in function
{\tt FT3SD5} each $X\in L$ is examined only once and one sole edge is added to each vertex in $\bigcup X.M$ (see formation
of set $E_U$).

Let $x\in\bigcup_{X\in L}X.U$. So, there is an $X_x\in L$, such that $x\in X_x.U$. Assume there is another $Y_x\in L$, such that
$x\in Y_x.U$.  First, assume that $X_x\in{\cal C}^u$ and $Y_x\in{\cal C}^v$. But $L$
is a clique of $\Gamma$, so edge $X_xY_x$ is in $\Gamma$. By construction of edge set of $\Gamma$ (figure~\ref{f:graph}),
$X_x.U\cap Y_x.U=\emptyset$, a contradiction. Second, assuming that $X_x\in{\cal C}^v$ and
$Y_x\in{\cal C}^u$ similarly leads to a contradiction. Third, assume that both $X_x$ and $Y_x$ belong to ${\cal C}^u$.
To form $X_x$ function {\tt Get\_structure} (figure~\ref{f:structures}) is called with input ($G, u, {\cal Q}, Q$), where
$Q$ is some component in ${\cal Q}$. Also, $Y_x$ must be formed by calling function {\tt Get\_structure} with
input ($G, u, {\cal Q}, Q'$), where $Q'$ is some component in ${\cal Q}$. Both of $Q$ and $Q'$ must be in the component of $G\setminus N_G[u]$ that contains $x$, so $Q\in Y_x.M$ and $Q'\in X_x.M$. Without loss of generality, assume that structure
$X_x$ is formed first. Since $Q'\in X_x.M$, algorithm {\tt T3SD5} cannot pick $Q'$ in order to call function {\tt Get\_structure}
with input ($G, u, {\cal Q}, Q'$), a contradiction. Fourth, similarly, assuming that both $X_x$ and $Y_x$ belong to ${\cal C}^v$
leads to a contradiction. All four cases lead to a contradiction, therefore, for each $x\in\bigcup_{X\in L}X.U$ there is a
unique $X_x\in L$ such that $x\in X_x.U$.
Hence, each vertex in $\bigcup_{X\in L}X.U$ has degree 1, because in function {\tt FT3SD5} each $X\in L$ is examined only
once and one sole edge is added to each vertex in $X.U$ (see formation of set $E_U$).

Let $A=\bigcup_{X\in L}(X.U\cup\bigcup X.M)$. First, each vertex in $A$ is adjacent in $T$ to a vertex in $\bigcup_{X\in L}X.D$, by
formation of set $E_U$ in function {\tt FT3SD5}. Second, $A\cap\bigcup_{X\in L}X.D=\emptyset$; to see this consider a
$y\in Y.D$, where $Y\in L$. Without loss of generality, assume that $Y\in {\cal C}^u$. Towards a contradiction, assume that
there is a $Z\in L$, such that $y\in Z.U\cup\bigcup Z.M$. Here, $y$ cannot be in $\bigcup Z.M$, because $Z.M$ is a set of
components of $G\setminus N[u, v]$ and $y\in N_G(u)$ (see figure~\ref{f:structures}). So, $y$ must be in $Z.U$. On one
hand, assume that $Z\in{\cal C}^u$; then $Z.U$ is subset of a component of $G\setminus N_G[u]$ and $y\in N_G(u)$, a
contradiction. On the other hand, assume that $Z\in{\cal C}^u$; then, since $L$ is a clique, edge $YZ$ is in $\Gamma$. But
this means that $Y.U\cap Z.D=\emptyset$ (see formation of edge set in figure~\ref{f:graph}), a contradiction. Third,
as proved in the previous paragraphs, each vertex in $A$ is a pendant vertex of $T$. Let $\bar{A}=V(G)\setminus A$. Therefore,
from these three facts, in order to prove that $T$ is a tree it suffices to prove that $T[\bar{A}]$ is a tree.

Set $\bar{A}$ is partitioned in $\bar{A}\cap V_L$ and $\bar{A}\setminus V_L$, where $V_L$ is defined in function {\tt FT3SD5}.
On one hand, set $\bar{A}\cap V_L$ is equal to $\bigcup_{X\in L}X.D$, because $\bigcup_{X\in L}X.U\subseteq A$ and
$A\cap\bigcup_{X\in L}X.D=\emptyset$ (see definition of $A$ and second fact of previous paragraph). Then, by formation
of edge set $E_D$ in function {\tt FT3SD5}, each vertex in $\bar{A}\cap V_L$ can be adjacent only to $u$ or to $v$ in $T$.
Assume, towards a contradiction, that a vertex $w\in \bar{A}\cap V_L$ is adjacent to both $u$ and $v$ in $T$. Then, there
must be a $Y\in L$, such that $Y\in {\cal C}^u$ and $w\in Y.D$, because edge $uw\in T$. Also, there must be a $Z\in L$, such that
$Z\in {\cal C}^v$ and $w\in Z.D$, because edge $vw\in T$. But $L$ is a clique, so $YZ$ is an edge of $\Gamma$ and,
therefore, $Y.D\cap Z.D=\emptyset$ (see formation of edge set of $\Gamma$ in figure~\ref{f:graph}), a contradiction.
So, every vertex in $\bar{A}\cap V_L$ is a pendant vertex adjacent in $T[\bar{A}]$ to a vertex outside of $\bar{A}\cap V_L$.
Also, by the formation of edge set $E'$ in function {\tt FT3SD5}, it is easily seen that $T[\bar{A}\setminus V_L$] is a tree (note
that $\bar{A}\subseteq N_G[u,v]$). Therefore, $T[\bar{A}]$ is a tree, which makes $T$ also a tree.

Each vertex in $\bar{A}$ is adjacent in $T$ to $u$ or to $v$. Also, each vertex in $A$ is adjacent to a vertex in $\bar{A}$. So,
every vertex of $G$ is within distance 2 from $\{u,v\}$ in $T$. Therefore, $T$ has diameter at most 5; note that edge $uv$ is
in $T$ (see edge set $E'$ in function {\tt FT3SD5}).

Consider any edge of $G$. If both of its endpoints are in $\bar{A}$, then each of them is within distance 1 from $u$ or $v$ in
$T$; so, they are within distance 3 apart in $T$. Therefore, in order to prove that $T$ is a 3-spanner of $G$ it suffices to
examine edges with at least one endpoint in $A$. Let $w$ be a vertex in $A$. Then, there is a (unique) $Y\in L$, such that
$w\in Y.U\cup\bigcup Y.M$. By construction of structure $Y$ (see figure~\ref{f:structures}),
$N_G(w)\subseteq Y.D\cup Y.U\cup\bigcup Y.M$. All vertices in $Y.U\cup\bigcup Y.M$ (including $w$) are adjacent in $T$ to the
same vertex of $Y.D$ (see formation of edge set $E_U$ in function {\tt FT3SD5}). Also, all vertices in $Y.D$ are adjacent in $T$
to the same vertex, $u$ or $v$. So, each vertex in $N_G(w)$ is within distance 3 in $T$ from $w$.\myqed

\section{Conclusions}
Algorithm {\tt T3SD5} described in section~\ref{s:description} is clearly efficient. Let $n(G)=|V(G)|$ be a function from
the set of graphs to the non negative integers. Then, the {\tt for} loop of the algorithm, that examines each edge of the
input graph, is executed $O(n^2)$ times. At the beginning of each execution of this loop is the formation
of set ${\cal Q}$, which can be done easily in $O(n^2)$ time. Next, each {\tt do-while} loop is executed $O(n)$ times;
to see this consider the formation of set ${\cal C}^u$: at each execution of the corresponding loop $\bigcup_{X\in{\cal C}^u}X.M$
is increased by at least one element of ${\cal Q}$. Each call of function {\tt Get\_structure} takes $O(n)$ time, since the
set of components of $G\setminus N_G[x]$ can be computed only twice (once for $u$ and once for $v$) and before entering
each {\tt do-while} loop. So, the two {\tt do-while} loops take time $O(n^2)$. The creation of graph $\Gamma$ takes
$O(n^2)$ time. To see this note that $\Gamma$ is the complement of a bipartite graph, so the time consuming operation
is to compute edges between ${\cal C}^u$ and ${\cal C}^v$. For this, examine each vertex of $G$ and if it belongs to two
structures of different sides don't place the edge between them; place all the remaining edges. Finally, function {\tt Check\_clique}
takes time $O(n^2)$, because its {\tt while} loop is executed $O(n)$ times and the commands within the loop take $O(n)$ time.
Particularly, to construct set $K$ in linear time one has to build a correspondence between elements of ${\cal Q}$ and structures in
${\cal C}^v$ before the execution of the {\tt while} loop so as to find in which structure each element of ${\cal Q}$ belongs to
in constant time. Therefore, the time complexity of the algorithm is $O(n^4)$.

Let $G$ be a graph and $uv$ one of its edges. Based on the description of function {\tt FT3SD5} in figure~\ref{f:function},
one can produce all $uv$-concentrated tree 3-spanners of $G$. Indeed, as shown in the proof of correctness,
each such spanner, corresponds to a choice of clique of structures $L$ (note that $L=\emptyset$, when ${\cal Q}=\emptyset$)
and to a choice of representative in each structure of $L$ (set $X.R$ is the set of representatives of structure $X$).
Also, each such spanner corresponds to a choice of edge set $E'$ in
function {\tt FT3SD5}, which has to do with neighbors of $u$ or $v$ that do not participate in any structure of $L$. Note that
there can be such spanners for which the corresponding set $E'$ contains edges that are not incident to either $u$ or $v$.
There are no other parameters to built such a spanner; so, all $uv$-concentrated tree 3-spanners of $G$ are listed
by choosing clique, representative, and set $E'$. To build all the tree 3-spanners of $G$, one can start with
$xy$-concentrated tree 3-spanners of $G$ for each edge $xy$ of $G$ and then try alter only the down vertex
adjacencies and set $E'$. The down vertices for each $xy$-concentrated tree 3-spanner $T$ of $G$ form set
$\bigcup_{X\in L_T}X.D$, where $L_T$ is the structure clique that corresponds to $T$. 

Small diameter graphs admit small diameter tree spanners:
\begin{lemma}
\label{l:graphdiam}
Let $G$ be a graph of diameter at most 3. Then, if $G$ admits a tree 3-spanner, then $G$ admits a tree
3-spanner of diameter at most 5.
\end{lemma}

\begin{figure}[htbp]
\begin{center}
\includegraphics[width=.7\textwidth]{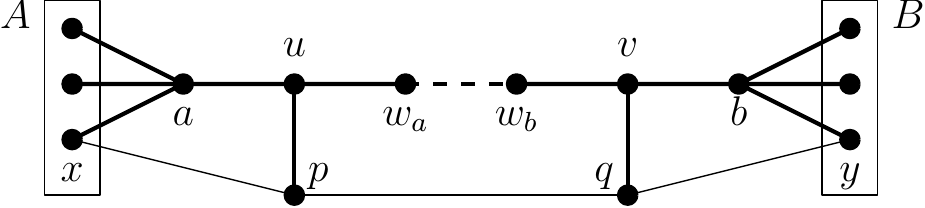}
\caption{Here, $x$, $y$ are endpoints of a diameter of $T$. The dashed line corresponds to a $T$ path of length at least 0; i.e.
it may be that $w_a=w_b$. It may also be $a=p$ or $b=q$.}
\label{f:graphdiam}
\end{center}
\end{figure}

$Proof$. Consider the tree 3-spanners of $G$ of smallest diameter. Among these, let $T$ be one that has the least
number of vertex pairs at $T$ distance equal to its diameter apart. Assume, towards a contradiction, that the diameter
of $T$ is strictly greater than 5. Let $D$ be a diameter of $T$. Let $a$ be the second vertex of path $D$ and
$b$ be the second last vertex of path $D$. Clearly, $a\not=b$. Also, let $u$ be the $D$ neighbor of $a$ towards $b$ and
let $v$ be the $D$ neighbor of $b$ towards $a$. Again, since the diameter is at least 6, $u\not=v$. Finally, let $w_a$ be
the $D$ neighbor of $u$ towards $b$ and let $w_b$ be the $D$ neighbor of $v$ towards $a$. Again, $w_a\not=v$ and $w_b\not=u$. Though, it can be the case that $w_a=w_b$ (see figure~\ref{f:graphdiam}).

Let $A$ be the $T$ neighbors of $a$ that are leaves and let $B$ be the $T$ neighbors of $b$ that are leaves.
All $T$ neighbors of $a$ but $u$ must be leaves, because $a$ is a second vertex of $D$. Also, all $T$ neighbors of
$b$ but $v$ must be leaves. Assume that a vertex $x\in A$ is adjacent to $u$ in $G$. Since $T$ is a tree 3-spanner
of $G$, $N_G(x)\subseteq A\cup N_T[u]$. So, tree $T'=(T\setminus\{xa\})\cup\{xu\}$ is a tree 3-spanner of $G$. But
in the new tree $x$ is moved closer to a center, so $T'$ has fewer than $T$ vertex pairs at distance equal to its diameter apart;
a contradiction. So $u\not\in N_G(A)$. Similarly, $v\not\in N_G(B)$.

Assume that some $x\in A$ is not adjacent to $w_a$ in $G$ and that some $y\in B$ is not adjacent to $w_b$ in $G$.
Since $G$ has diameter at most 3, there is a path $x,p,q,y$ in $G$, where $p$ can be equal to $q$; note that $x,y$
are too far apart in $T$ to be adjacent in $G$. Since $T$ is a tree 3-spanner of $G$, $p$ must be in
$(A\cup N_T(u))\setminus\{w_a\}$;
note that $x$ is not adjacent in $G$ to either of $u$ or $w_a$. Similarly, $q$ must be in $(B\cup N_T(v))\setminus\{w_b\}$.
If $p\in A$, then $p$ cannot be a neighbor of $q$, because, then, the closest to $y$ possibly neighbor of $p$ is $w_a$
and $q$ cannot be equal to $w_a$, even if $w_a=w_b$. So, $p$ must be in $N_T(u)\setminus\{w_a\}$. Similarly,
$q$ must be in $N_T(v)\setminus\{w_b\}$ (see figure~\ref{f:graphdiam}).
Then, $d_T(p,q)=d_T(p,u)+d_T(u,v)+d_T(v,q)$. But $d_T(u,v)\geq 2$;
so, this is a contradiction, because $p$ and $q$ are adjacent in $G$ and $T$ is a tree 3-spanner of $G$. Therefore,
all the vertices of $A$ are adjacent in $G$ to $w_a$ or all the vertices of $B$ are adjacent in $G$ to $w_b$.

So, without loss of generality, all vertices of $A$ are adjacent in $G$ to $w_a$. Let $T'$ be the
tree $(T\setminus\bigcup_{z\in A}\{za\})\cup\bigcup_{z\in A}\{zw_a\}$. Then, $T'$ is a tree 3-spanner of $G$. But
in the new tree $x$ is moved closer to a center, so $T'$ has fewer than $T$ vertex pairs at distance equal to its diameter apart;
a contradiction. \myqed

\begin{thm}
\label{t:graphdiam}
There is an efficeint algorithm to decide whether a graph of diameter at most 3 admits a tree 3-spanner.
\end{thm}
$Proof$. Based on lemma~\ref{l:graphdiam} and theorem~\ref{t:final} algorithm {\tt T3SD5($G$)} in figure~\ref{f:algorithm}
decides tree 3-spanner admissibility of diameter at most 3 graphs. As noted earlier in this section this algorithm is efficient.\myqed

Note that a graph $G$ admits a tree 3-spanner if and only if every block of $G$ admits a tree 3-spanner. So,
algorithm {\tt T3SD5($G$)} can be employed to decide tree 3-spanner admissibility of bigger diameter graphs, as
long as each block of the input graph has diameter at most 3.

A long standing open question is to determine the complexity status of the tree 3-spanner problem: given a graph
decide whether it admits a tree 3-spanner, without any diameter restrictions. It seems that the algorithm
presented in this paper can be used as a building block towards deciding tree 3-spanner admissible graphs. First, relaxing
the diameter restriction and, therefore, finding tree 3-spanners of longer diameter, eventually the problem will be solved
for any diameter and will cover all tree 3-spanner admissible graphs. Second, tree 3-spanner admissible graphs of diameter
more than 3 should admit a star cut-set; i.e. a cut-set consisting of a vertex and some of its neighbors. So, it may be the
case that deciding whether a graph $G$ admits a tree 3-spanner is reduced to deciding the tree 3-spanner admissibility
of a set of small diameter subgraphs of $G$.

\bibliographystyle{plain}
\bibliography{tspanners}
\end{document}